\documentclass[11pt]{article}
\usepackage{jhep}
\usepackage[T1]{fontenc}
\usepackage{bm}
\usepackage{graphicx}
\usepackage{caption}
\usepackage{subcaption}
\usepackage[normalem]{ulem}
\usepackage{gensymb}

\newcommand{\beq}{\begin{equation}} 
\newcommand{\eeq}{\end{equation}}  
\newcommand{\bea}{\begin{eqnarray}}  
\newcommand{\eea}{\end{eqnarray}}

\newcommand{\GeV}{\text{GeV}}

\newcommand{\lh}{\lambda_{h}}

\definecolor{GW}{rgb}{1,0,0}
\definecolor{DM}{rgb}{1,0,1}

\title{Stability, reheating, and leptogenesis}
\author[a]{Djuna Croon,}
\author[b,c]{Nicolas Fernandez,}
\author[a]{David McKeen}
\author[a]{and Graham White}

\affiliation[a]{TRIUMF, 4004 Wesbrook Mall, Vancouver, BC V6T 2A3, Canada}
\affiliation[b]{Department of Physics, 1156 High St., University of California Santa Cruz, Santa Cruz, CA 95064, USA}
\affiliation[c]{Santa Cruz Institute for Particle Physics, 1156 High St., Santa Cruz, CA 95064, USA}
\emailAdd{dcroon@triumf.ca}
\emailAdd{nfernan2@ucsc.edu}
\emailAdd{mckeen@triumf.ca}
\emailAdd{gwhite@triumf.ca}
\date{\today}
\abstract{
In a minimal model of leptogenesis, the observed baryon asymmetry is realized after high-scale reheating into the lightest sterile neutrino. 
We consider constraints on this scenario from the stability of the Higgs vacuum during pre-heating. 
Depending on the reheat temperature, the lightest sterile neutrino may be in or out of thermal equilibrium at production. Demanding stability of the Higgs vacuum during pre-heating, we find strong constraints   which primarily impact the parameter space of thermal leptogenesis.}

\begin{document}
\maketitle

\section{Introduction}
Leptogenesis~\cite{Fukugita:1986hr} is an attractive explanation of the baryon asymmetry of the Universe (BAU). Traditional models of leptogenesis, which we term ``vanilla,'' involve CP violating decays of heavy sterile neutrinos, resulting in a lepton asymmetry which is transferred to baryons through electroweak sphalerons. For reviews, see, e.g.~\cite{Buchmuller:2004nz,Buchmuller:2005eh,DiBari:2012fz,Blanchet:2012bk,White:2016nbo}. Such models may simultaneously explain the light neutrino masses via the seesaw mechanism~\cite{Minkowski:1977sc,*Yanagida:1979as,*GellMann:1980vs,*Glashow:1979nm,*Mohapatra:1979ia,*Schechter:1980gr}.
A minimal model capable of explaining current neutrino oscillation data and allowing for leptogenesis extends the standard model (SM) particle content with just two sterile neutrinos with masses $M_{1,2} \gtrsim 10^{10}$ GeV.  

Alas, the large sterile neutrino masses make vanilla leptogenesis notoriously difficult to probe experimentally. Of the eleven free parameters in a neutrino sector with two steriles and normal mass ordering, a global fit of neutrino oscillation data allows one to determine six: two mass splittings in the SM neutrino sector, three mixing angles, and the Dirac phase in the Pontecorvo-Maki-Nakagawa-Sakata matrix (PMNS matrix) \cite{Esteban:2018azc,nufit}. Convincing evidence for the vanilla leptogenesis paradigm then relies on indirect experimental access to the Majorana phase or the masses of the heavy sterile neutrinos which would certainly be hard to come by.

In this paper, we propose stringent cosmological constraints on the vanilla leptogenesis paradigm from stability of the Higgs vacuum during preheating. We argue that a relic abundance of heavy sterile neutrinos---necessary to produce a large lepton asymmetry---strongly motivates neutrino production during reheating through a coupling to the inflaton. In the vanilla leptogenesis model, such a coupling generates Higgs-inflaton couplings at the 1-loop level. This coupling, in turn, may lead to catastrophic decay of the Higgs vacuum; in the SM, the presently measured central values of the Higgs and top masses along with the strength of the strong gauge coupling lead to a negative Higgs quartic at a scale around $10^{10}~\rm GeV$. As such, this scenario is subject to very strong constraints~\cite{Enqvist:2016mqj,Ema:2016kpf,Ema:2017loe}.\footnote{The stability of the Higgs vacuum has also been considered in the context of inflation, e.g.,~\cite{Hook:2014uia,Kearney:2015vba,Shkerin:2015exa,Herranen:2014cua,Espinosa:2017sgp}, and reheating, e.g.,~\cite{Rose:2015lna,Enqvist:2016mqj,Ema:2017ckf,Figueroa:2017slm}.} 

As we will see, demanding stability of the Higgs vacuum during preheating introduces strong constraints on the Yukawa matrices in the neutrino sector. This translates into constraints on the mass of the lightest sterile neutrino and significantly reduces the available parameter space for vanilla leptogenesis. Our results are complementary to the zero temperature stability implications~\cite{Bambhaniya:2016rbb,Ipek:2018sai}, which provide an upper bound on the Yukawa couplings between the Higgs and the neutrinos. 

To demonstrate our constraints, we consider the most conservative scenario, in which the inflaton only couples to the lightest sterile neutrino. As any coupling to heavier neutrinos induces a larger one-loop Higgs-inflaton coupling, the constraints on such scenarios are expected to be much stronger. 
We study the scenario in which the lightest sterile is produced in thermal equilibrium ($M_1\lesssim T_R$), as well as out of equilibrium ($M_1\gtrsim T_R$). We find that our constraints are most important in the thermal leptogenesis regime, and can thus be interpreted as an argument for non-thermal leptogenesis (beyond the usual arguments of (i) its high efficiency, since washout from inverse decays are suppressed~\cite{Buchmuller:2004nz,Antusch:2010mv,Antusch:2018zvu}, and (ii) allowing for leptogenesis without producing relics that disturb cosmological successes such as gravitinos in supersymmetry~\cite{Asaka:2000zh}).

The structure of this paper is as follows. We summarize our vanilla model in section~\ref{sec:model}. We give a brief review of preheating and motivate the constraints for a range of inflaton masses in section~\ref{sec:reheating} and apply them to both thermal and non-thermal leptogenesis while accounting for neutrino oscillation data in sections~\ref{sec:leptogenesis} to~\ref{sec:results}. Finally we conclude with a discussion in section~\ref{sec:discussion}.

\section{Vanilla leptogenesis}\label{sec:model}
\subsection{Model}
A minimal model that can explain the currently observed neutrino masses, the BAU, and inflation extends the SM Lagrangian with two sterile neutrinos, $N_{1,2}$, and an inflaton, $\phi$. The Lagrangian is
\begin{equation}
\begin{aligned}
    \mathcal{L} &= \mathcal{L}_{SM}+\frac{1}{2}(\partial \phi)^2- V(\phi )-\frac{g}{2}\, \phi^2 |H|^2 -\sigma \phi |H|^2 \\
    &\quad+i\bar N_i\!\!\not\!\partial N_i- \left(\frac12 M_{ij}\bar{N}_i^c N_j+ Y_{ij}\bar L_i \tilde H  N_j+\tilde{ \lambda } _{ij} \phi\bar{N}_i^c N_j+ {\rm h.c.}\right),
\end{aligned}
\label{eq:model}
\end{equation}
with $L$ ($H$) the lepton (Higgs) doublet and $\tilde H=i\sigma^2H^\ast$. We will work in a basis where the sterile neutrino mass matrix is diagonal, $M={\rm diag}(M_1,M_2)$. 


In this basis, the constraints are weakest when the inflaton couples dominantly to the lightest sterile neutrino, $\tilde\lambda={\rm diag}(\lambda,0)$, which is the case we will focus on below; coupling of the inflaton to the heavier sterile neutrino is more strongly constrained. Because of the larger $N_2$ Yukawa, a $\phi$-$N_2$-$N_2$ coupling leads to larger contributions to the induced inflaton-Higgs coupling resulting in tighter constraints from the requirement of stability during preheating. Thus, this pattern of couplings and masses leads to conservative bounds. Building a model that generates such couplings naturally, while interesting, is beyond the scope of this work.

Integrating out the heavy sterile neutrinos generates a nonzero mass matrix for the light neutrinos,
\begin{equation}
\tilde m_\nu=-v^2YM^{-1}Y^T 
\end{equation}
where $v=174~\rm GeV$ is the Higgs field vev. This is diagonalized by the PMNS matrix, $U$: $U^T \tilde m_\nu U=m_\nu={\rm diag}(m_1,m_2,m_3)$, with $U$ a unitary matrix described by three mixing angles, $\theta_{12}$, $\theta_{23}$, and $\theta_{13}$, a Dirac phase $\delta$ and two Majorana phases $\alpha_{21}$ and $\alpha_{31}$. We assume a normal hierarchy and work in the minimal setup that can explain neutrino oscillations, with $m_1=0$, $m_2=\sqrt{\Delta m_\odot^2}=(8.6\pm0.1)\times10^{-3}~\rm eV$, and $m_3=\sqrt{\Delta m_{\rm atm}^2}=(5.02\pm0.03)\times10^{-2}~\rm eV$ where $\Delta m_\odot^2$ and $\Delta m_{\rm atm}^2$ are the solar and atmospheric mass splittings, respectively, and $31.6\degree<\theta_{12}<36.3\degree$, $40.9\degree<\theta_{23}<52.2\degree$, and $8.22\degree<\theta_{13}<8.98\degree$ at $3\sigma$~\cite{Esteban:2018azc,nufit}. The Dirac and Majorana phases are so far unconstrained.

The Yukawa matrix $Y$ can be related to the physically measured masses and mixings using the Casas-Ibarra parameterization~\cite{Casas:2001sr}
\begin{equation}
Y=\frac{1}{v}U\sqrt{m_\nu}R^T\sqrt{M},
\end{equation}
where $R$ is an undetermined orthogonal matrix and the heavy and light mass matrices, $m_\nu$ and $M$, have been diagonalized. In the case of two sterile neutrinos, the orthogonal matrix takes the simplified form~\cite{Ibarra:2003up,Antusch:2011nz}
\begin{equation}
R=\left(\begin{array}{ccc}
0 & \cos z & \pm\sin z \\
0 & -\sin z & \pm\cos z \\
1 & 0 & 0
\end{array}\right),
\end{equation}
with $z$ a complex angle. Thus, our neutrino sector contains five parameters fixed by experiment, $\theta_{12}$, $\theta_{23}$, $\theta_{13}$, $m_2$, and $m_3$ and six unconstrained parameters, $\delta$, $\alpha_{21}-\alpha_{31}$, $M_1$, $M_2$, ${\rm Re}\,z$, and ${\rm Im}\,z$ (our choice of two sterile neutrinos fixes $m_1=0$ and renders $\alpha_{21}+\alpha_{31}$ unphysical).

\begin{figure}
    \centering
    \includegraphics[width=0.65\textwidth]{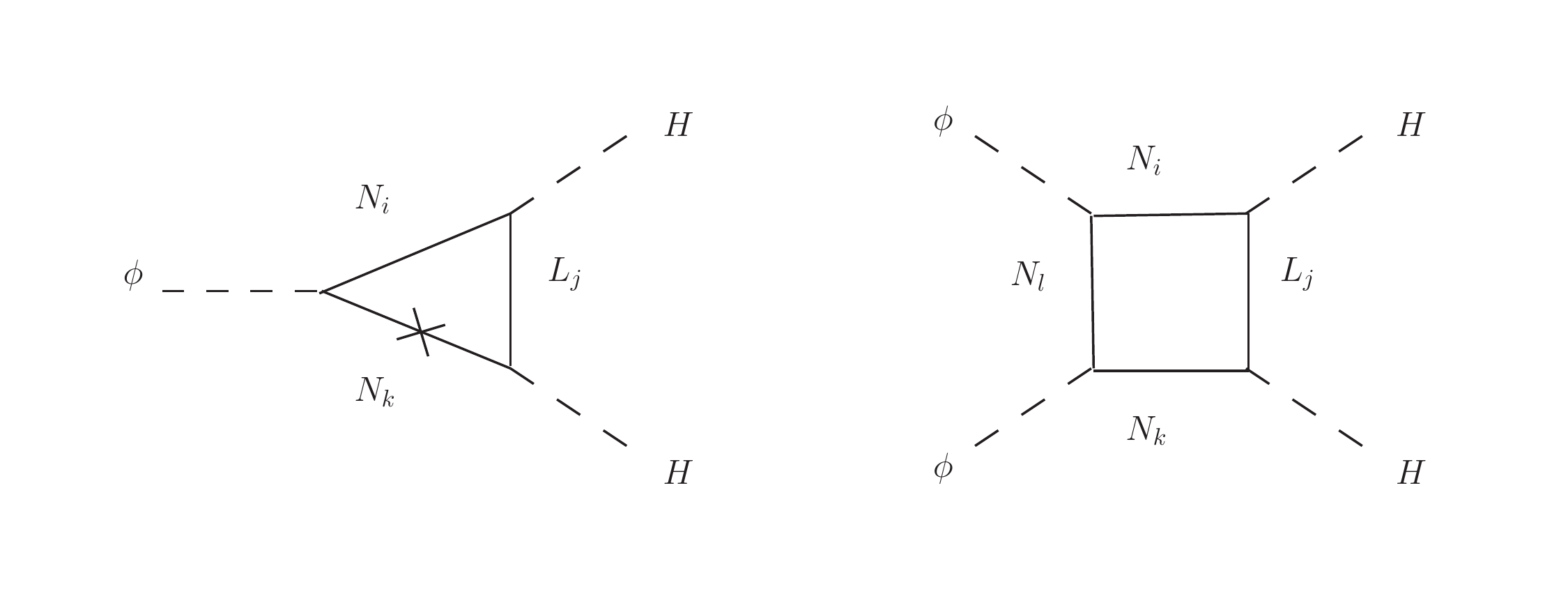}
    \caption{Radiatively generated Higgs-inflaton couplings.}
    \label{fig:portals}
\end{figure}

Even if defined to vanish at some scale, the portal couplings $g$ and $\sigma$ in Eq.~(\ref{eq:model}) obtain nonzero values at other scales due to radiative effects. The leading contributions come at one-loop with virtual sterile neutrinos as shown in Fig.~\ref{fig:portals}.
The shifts of the couplings are
\begin{align} 
\delta\sigma &=-\frac{1}{4\pi^2}{\rm Tr}\left[{\rm Re}\left(\tilde\lambda M+M\tilde\lambda\right)Y^\dagger Y\right] \log \frac{\Lambda}{m_\phi}=-\frac{\lambda}{2\pi^2} M_1 (Y^\dag Y)_{11}\log\frac{\Lambda}{m_\phi}, \\
\delta g &= \frac{1}{2\pi^2}{\rm Tr}\left[\tilde\lambda^\dagger \tilde\lambda Y^\dagger Y\right] \log \frac{\Lambda}{m_\phi}=\frac{\lambda^2}{2\pi^2} (Y^\dag Y)_{11}\log\frac{\Lambda}{m_\phi},\label{eq:portals}
\end{align}
where we have specialized to our basis with $\tilde\lambda={\rm diag}(\lambda,0)$ on the right-hand sides of these equations. In these expressions, $\Lambda$ is the renormalization scale at which the portal couplings are defined. Following~\cite{Enqvist:2016mqj}, we take this scale to $\Lambda=M_{\rm Pl}$ (where $M_{\rm Pl}=1.2\times10^{19}~\rm GeV$ is the Planck mass). Since the shifts only depend logarithmically on this scale, our results are not very sensitive to this particular choice. We will use these shifts as estimates of the minimum size of the portal couplings in a natural theory,
\begin{equation}
\left|g\right|\gtrsim \left|\delta g\right|,~\left|\sigma\right|\gtrsim \left|\delta \sigma\right|.
\end{equation}
Stated differently, this is essentially equivalent to assuming that the portal couplings are defined to vanish at the scale $\Lambda=M_{\rm Pl}$.

The quartic coupling of the Higgs in $\mathcal{L}_{SM}\supset\lambda_H\left|H\right|^4$ also runs due to quantum corrections. The dominant source of its running comes from the coupling of the Higgs to the top quark, which drives $\lambda_H$ toward negative values. This can destabilize the Higgs potential at large field values, which will constrain the allowed parameter space in this model. Current measurements of the masses of the top quark and the Higgs boson as well as the value of the strong coupling constant indicate that the electroweak vacuum may have a catastrophic vacuum at high scale, with the instability scale between $10^{8}$~GeV and $10^{16}$~GeV \cite{Markkanen:2018pdo,Andreassen:2014gha}. For our analysis, we will adopt the central value of $10^{10}$~GeV, noting that the results do not depend strongly on this particular value. 

As for the inflationary portion of our model in~\ref{eq:model}, our results apply in principle to any inflationary potential that is  approximately quadratic near $V=0$. Besides chaotic inflation, this includes, for example, models in which the inflaton is a pseudo-Goldstone boson \cite{Croon:2014dma,Croon:2015fza,Croon:2015naa}, and alpha-attractor potentials \cite{Kallosh:2013hoa,Kallosh:2013yoa}. 
In these models, a symmetry protects the flatness of the potential, such that quantum corrections do not spoil slow roll.\footnote{In such models the coupling $\tilde{ \lambda } _{ij} \phi\bar{N}_i^c N_j$ can be considered the leading term in an expansion.} In the absence of such a symmetry, quantum corrections may ruin the flatness of the inflaton potential, implying an upper bound on the coupling $\tilde\lambda_{ij}$ in \eqref{eq:model}. In a seesaw model of neutrino masses, this leads to an upper bound on $M_1$, as was previously considered in \cite{Enqvist:2016mqj}. 

\subsection{Baryon asymmetry}\label{sec:leptogenesis}
We briefly outline here the calculation of the baryon asymmetry in this model, which proceeds through the lepton-number-violating asymmetry in the decay of the sterile neutrinos. As we will see below, production of the sterile neutrinos is dominated by perturbative decay of the inflaton. As mentioned above, we conservatively assume that the inflaton coupling to the sterile neutrinos is dominantly to the lightest one. Furthermore, we will take the sterile neutrino masses to be hierarchical, and hence $N_2$ is irrelevant for the asymmetry that is produced. In this case, the equations that govern the evolution of the energy densities in the inflaton, $\rho_\phi$, sterile neutrino $N_1$, $\rho_{N_1}$, and radiation, $\rho_R$, are\footnote{Here, we neglect flavour effects. For a thorough treatment of such effects see~\cite{Moffat:2018wke}.}
\begin{equation}
\begin{aligned}
\frac{d\rho_\phi }{dt}+3H\rho_\phi &= -\Gamma_\phi\,  \rho_\phi, \\
\frac{d\rho_{N_1}}{dt}+3H\rho_{N_1} &= -\Gamma_{N_1}\left(\rho_{N_1}-\rho_{N_1}^{\rm eq}\right) + \Gamma_{\phi\to N_1 N_1}\,  \rho_\phi, \\
\frac{d\rho_R}{dt}+4H\rho_R &= \Gamma_{N_1}\left(\rho_{N_1}-\rho_{N_1}^{\rm eq}\right) + \Gamma_{\phi\to R}\,  \rho_\phi.
\end{aligned}
\label{eq:Boltzmann}
\end{equation}
In these expressions, the total inflaton width, $\Gamma_\phi=\Gamma_{\phi\to R}+\sum_i \Gamma_{\phi\to N_i N_i}=\Gamma_{\phi\to R}+\Gamma_{\phi\to N_1 N_1}$, is written in terms of the partial widths to radiation (through its coupling to the Higgs) and sterile neutrinos respectively,
\begin{equation}
\Gamma_{\phi\to R}=\frac{1}{16\pi}\frac{\sigma^2}{m_\phi},~\Gamma_{\phi\to N_1 N_1}=\frac{\lambda^2}{8\pi}m_\phi\left(1-\frac{4M_1^2}{m_\phi^2}\right)^{3/2},
\label{eq:phidecayrate}
\end{equation}
and the thermally averaged sterile neutrino width is
\begin{equation}
\Gamma_{N_1}=\frac{K_1(M_1/T)}{K_2(M_1/T)}\frac{(Y^\dagger Y)_{11}}{8\pi}M_1,
\end{equation}
where $K_{1,2}$ are modified Bessel functions of the second kind. The temperature is related to the radiation energy density, $\rho_R=g_\ast \pi^2 T^4/30$ where $g_\ast\simeq106.75$ counts the relativistic degrees of freedom during the time in question. $\rho_{N_1}^{\rm eq}$ is the equilibrium $N_1$ energy density which depends on their temperature.

The sterile neutrinos produce a lepton number asymmetry that is quickly turned into an asymmetry in $B-L$, the difference between baryon and lepton number, by electroweak sphalerons. Once Eqs.~(\ref{eq:Boltzmann}) are solved, the $B-L$ number density, $n_{B-L}$, can be found through
\begin{equation}
\frac{dn_{B-L}}{dt}+3Hn_{B-L}= \frac{\epsilon_1\Gamma_{N_1}}{M_i}\left(\rho_{N_1}-\rho_{N_1}^{\rm eq}\right)-W n_{B-L}.
\end{equation}
Since we will be working with a hierarchical sterile neutrino spectrum, $M_2\gg M_1$, the asymmetry is dominated by $N_1$ decays. The asymmetry parameter is~\cite{Covi:1996wh,Abada:2006fw,DeSimone:2006nrs,Blanchet:2011xq}
\begin{equation}
\epsilon_1=\frac{3\,{\rm Im}\left[(Y^\dagger Y)_{21}^2\right]}{16\pi (Y^\dagger Y)_{11}}f\left(\frac{M_2^2}{M_1^2}\right),~f(x)\equiv\frac{2\sqrt x}{3}\left[\left(1+x\right)\log\left(\frac{1+x}{x}\right)-\frac{2-x}{1-x}\right].
\end{equation}
The washout term is related to the $N_1$ decay rate which parameterizes the strength of $N_1$'s coupling to the plasma,
\begin{equation}
W=\frac12 \Gamma_{N_1}\frac{n_{N_1}^{\rm eq}}{n_{B-L}^{\rm eq}},
\end{equation}
with $n_{N_1}^{\rm eq}$ the equilibrium number density of $N_1$. The relevant terms involving the Yukawa matrix here can be expressed in terms of the parameterization above as
\begin{equation}
(Y^\dagger Y)_{ij}=\frac{\sqrt{M_iM_j}}{v}\left(\frac{m_3}{v}R^\ast_{i3}R_{j3}+\frac{m_2}{v}R^\ast_{i2}R_{j2}\right),
\end{equation}
or in particular,
\begin{equation}
\begin{aligned}
(Y^\dagger Y)_{11}&=\frac{M_1}{v}\left(\frac{m_3}{v}\left|\sin z\right|^2+\frac{m_2}{v}\left|\cos z\right|^2\right),\\
(Y^\dagger Y)_{21}&=\frac{\sqrt{M_1M_2}}{v}\left(\frac{m_3}{v}\cos z^\ast\sin z-\frac{m_2}{v}\cos z\sin z^\ast\right).
\end{aligned}
\label{eq:YY11}
\end{equation}
Assuming the dominance of the terms proportional to $m_3$ in $(Y^\dagger Y)_{ij}$, the asymmetry parameter can be expressed as
\begin{equation}
\epsilon_1\simeq\frac{3}{16\pi}\frac{M_1 m_3}{v^2}\sin2\beta\left[1+\frac{5 M_1^3}{9M_2^3}+{\cal O}\left(\frac{ M_1^5}{M_2^5}\right)\right]\simeq 10^{-5}\sin 2\beta\left(\frac{M_1}{10^{11}~\rm GeV}\right), \label{eqn:eps}
\end{equation}
with $\beta=\arg\sin z$. The proportionality of $\epsilon_1$ to $M_1$ leads to the well-known Davidson{-}Ibarra lower bound on $M_1$~\cite{Davidson:2002qv}.  Note that in the hierarchical regime $M_1\ll M_2$ the analogous asymmetry from decays of $N_2$, $\epsilon_2$, is suppressed compared to $\epsilon_1$ by a factor of $M_1/M_2$.

The present day baryon-to-photon ratio can be found from the $B-L$ number density through
\begin{equation}
\eta_B=\frac{n_{B}}{n_\gamma}=\left(\frac{s}{n_\gamma}\right)_0\frac{n_B}{n_{B-L}}\frac{n_{B-L}}{s}=7.04\times\frac{28}{79}\times\frac{n_{B-L}}{s}
\end{equation}
where $s=2\pi^2g_\ast T^3/45$ is the entropy density, the ``0'' subscript indicates a quantity evaluated today, and the factor of $28/79$ relates the final $B$ asymmetry to the $B-L$ asymmetry generated above the electroweak scale.  We note that the baryon asymmetry can be characterized by the asymmetry parameter, $\epsilon _1$, and an ``efficiency parameter'', $\kappa_f$~\cite{Buchmuller:2004nz}, such that
\begin{equation}
    \eta _B \sim 10^{-2} \epsilon_1 \kappa_f.
\end{equation}
The efficiency parameter can be approximated as~\cite{Buchmuller:2004nz}
\begin{equation}
    \kappa _f \sim (2\pm 1)\times 10^{-2} \left( \frac{0.01~{\rm eV}}{\tilde{m}_1} \right)^{1.1} 
\end{equation}
with $\tilde{m}_1=v^2(Y^\dagger Y)_{11}/M_1$. Using \eqref{eq:YY11}, one finds that $\tilde{m}_1 \geq m_2\simeq0.0086~{\rm eV}$. This allows us to derive a bound for the efficiency parameter
\begin{equation}
   \kappa _f  \leq (2 \pm 1)\times 10^{-2} \left( \frac{0.01~{\rm eV}}{0.0086~ {\rm eV}} \right)^{1.1} \ ,
\end{equation}
from which we can derive a lower bound on $M_1$ of $10^{10}-10^{11}$ GeV. Note that the strictness of this bound arises from the minimality of our model. Relaxing the condition that $|\sin z|\leq 1$ as well as including flavour effects may relax this lower bound on $M_1$ \cite{Hambye:2003rt,Blanchet:2012bk}. A detailed analysis of this is beyond the scope of this work.

\section{Higgs instability during preheating}\label{sec:reheating}
After inflation, the inflaton field may decay to other fields perturbatively, or through a non-perturbative preheating mechanism. The latter process can produce high-energy Higgs modes which destabilize the Higgs vacuum. In this section, we will estimate the upper bounds on the portal couplings for which the Higgs vacuum is destabilized during the non-perturbative regime. 
As we will see below, for portal couplings below these bounds, the dominant mechanism through which energy is transferred from the inflaton field to the other fields is the perturbative decay.

\subsection{Higgs preheating}
To get a feel for the results of our lattice study, we first present an approximate analytic description of the preheating dynamics of the inflaton and Higgs fields.
This analytic description was developed in the early 90s~\cite{Traschen:1990sw,Kofman:1994rk,Shtanov:1994ce,Kofman:1997yn}; for a more recent review see, e.g.,~\cite{Allahverdi:2010xz}.
This section is meant to give a qualitative description of the parameter space that is allowed by Higgs stability during preheating. 

Slow-roll inflation ends when the inflaton field rolls to the minimum of its potential, where it starts a damped oscillation with friction given by the expanding background and couplings to other particles. 
For generic models, the effective potential of the inflaton field is approximately quadratic around this minimum. In our case, we also consider renormalizable couplings of the inflaton to the Higgs field as in \eqref{eq:model}, and therefore our scalar potential is
\beq
V(\phi,h)=\dfrac{1}{2}m_\phi^{2}\phi^{2} + \dfrac{g}{4}\phi^{2}h^{2} + \dfrac{\sigma}{2}\phi h^{2} + \dfrac{\lh}{4}h^{4}\,.
\eeq
which we expressed in terms of the radial component of the Higgs field $|H| = h/\sqrt{2}$. Here $m_\phi$ is the effective mass of the inflaton while it oscillates around the minimum of its potential.

During the oscillatory phase, the inflaton field has equation of state parameter $w = (n-1)/(n+1)$ for effective potential $V \sim \phi^{2n}$. Thus, for an approximately quadratic potential, the field behaves like a pressure-less fluid. In this regime one treats the inflaton field as an oscillating background, which solves
\begin{eqnarray}
\ddot{\phi} + 3 H \dot{\phi} + V'(\phi) =0. 
\end{eqnarray}
Here we assume that the inflaton dominates the energy density of the Universe and neglect backreaction, so that we can use $a\propto t^{2/3}$.
Assuming further that preheating does not complete for several oscillations of the inflaton field, the dynamics at late times ($m_\phi t\gg1$) can be approximated by 
\beq
\phi(t)= \Phi(t)\cos(m_\phi t)\,, \qquad \text{where} \qquad \Phi(t) \sim \dfrac{\Phi_0}{ m_\phi t}\,, \label{eq:inflosc}
\eeq
and where the value of $\Phi_0$ depends on the particular inflation model.\footnote{As the approximate description \eqref{eq:inflosc} only becomes accurate after the first few damped semi-oscillations ($mt\gg1$), $\Phi_0$ is typically smaller than the value of $\phi$ at the end of inflation. To compare to our lattice study, we will use $\Phi_{0} \approx 0.2 M_{\rm Pl}$.}  
In order to study the fluctuation and growth of modes of the Higgs field, we write the mode equation for Higgs field in the oscillating background \eqref{eq:inflosc}. 
In terms of the comoving field $\tilde{h}_{k}=a^{3/2} h_{k}$, 
\begin{eqnarray}
\label{eq:mathieu}
\ddot{\tilde{h}}_{k} + \omega_k^2 \tilde{h}_{k} = 0 \, , \,\,\,\,\,\,
\omega^{2}_{k} = \left[ \dfrac{k^{2}}{a^{2}} + \dfrac{g}{2} \Phi^{2} \cos^{2}(m_\phi t) + \sigma \Phi \cos(m_\phi t) + 3 a^{-3} \lh(h) \langle \tilde{h}^{2} \rangle + \Delta \right],
\end{eqnarray}
where we have used the Hartree approximation in the quartic term, and where {$\Delta=-9H^{2}/4-3\dot{H}/2$} which is negligible as long the equation of state is matter-like.
Here and in the following, we will assume preheating is fast so that we can set $a=1$ without loss of generality, and drop the tildes. 
For $\sigma=0$~\eqref{eq:mathieu} is known as the Mathieu equation, while for $\sigma \neq 0$ it is known as the Whittaker-Hill equation.
These equations have instabilities, and can in particular lead to exponential growth for certain wavenumbers $k$, 
\begin{equation}\label{eq:floquet}
    h_k \sim e^{\mu_k m_\phi t}
\end{equation} where $\mu_k>0$ is the Floquet exponent. Which wave numbers $k$ are sensitive to these instabilities depends on the behaviour of the resonance.
From \eqref{eq:mathieu} it is seen that there are two dimensionless parameters which control the amplification of the resonance,
\beq
q=\dfrac{g \Phi^{2}(t)}{2m_\phi^{2}} \qquad \text{and} \qquad p=\dfrac{2 \sigma \Phi(t)}{m_\phi^{2}}\,.
\eeq
For simplicity, we will first study the Mathieu equation, with  $p=0$, and we will also set the Higgs self-interaction to zero for the time being, $\lambda_h=0$. The behaviour of parametric resonance can then be classified as narrow for  $q \ll 1$ or broad for $q \gg 1$. In our case, the initial value of $q$ is 
\beq
q_{0} = \dfrac{g \, \Phi^{2}_{0}}{2m^{2}}  \approx 118 \left(\dfrac{g}{10^{-8}}\right) \left(\dfrac{1.3\times10^{-6} M_{\rm Pl}}{m_\phi}\right) \left(\dfrac{\Phi_{0}}{0.2 M_{\rm Pl}} \right)^{2}\,,
\eeq
where the right hand side is informed by our benchmark scenario, in which $g=10^{-8}$ and $m_\phi=1.3 \times10^{-6} M_{\rm Pl}$, as in chaotic inflation.
In the broad resonance regime, then, the Mathieu equation \eqref{eq:mathieu} has instabilities for large ranges of $k$.
The Higgs modes evolve adiabatically as long as the adiabaticity condition is satisfied, 
\begin{equation} \label{eq:adiabaticevolution}
    \dot{\omega_k^2} \leq 2 \omega_k^3 .
\end{equation}
Using \eqref{eq:mathieu}, it is seen that this is satisfied away from inflaton zero-crossings. In this regime, a WKB approximation can be used to solve the Mathieu equation \cite{Kofman:1997yn}, 
\begin{equation}
    h_k = \frac{1}{\sqrt{2\omega_k}} \left(\alpha_k e^{-i \int \omega_k dt}+\beta_k e^{i \int \omega_k dt} \right) .
\end{equation}
The occupation number of the Higgs field is then defined in terms of the second Bogolyubov coefficient, $n_k = |\beta_k|^2$. The variance of the Higgs field can be related to the mode occupation number as 
\begin{equation}
    \langle h^2 \rangle \approx \frac{1}{(2 \pi)^3} \int d^3k \, \left(\frac{n_k}{\omega_k} \right).
\end{equation}
In contrast, near the inflaton zero-crossings, $\omega_k$ evolves very quickly, and the adiabaticity condition \eqref{eq:adiabaticevolution} is violated. In this regime Higgs particles are created in violent bursts. 
It is found that the number density increases exponentially around zero-crossings $j$,
\begin{equation}
    n_k^{j+1} = n_k^j e^{2\pi\mu^j_k} 
\end{equation}
where $\mu_k$ is the Floquet exponent of the Mathieu equation, and zero crossings occur for integer multiples of $m_\phi t/\pi$. As a function of time, the Higgs variance can then be found to be \cite{Ema:2016kpf}
\begin{equation}\label{eq:higgsvariance}
    \langle h^2 \rangle \approx \frac{ \Phi(t)}{32 \pi^{3/2}} \sqrt{  \frac{g m_\phi}{ \mu t}} e^{\left(2 \mu m_\phi  - \Gamma_{h\rightarrow \bar{t}t}\right)t} .
\end{equation}
with Floquet exponent $\mu\approx (\ln3)/2\pi$ \cite{Allahverdi:2010xz}.
Here the second term in the exponential accounts for the perturbative decay of the Higgs into top-quarks, which slows down the growth of Higgs modes slightly.

\subsection{Parametric resonance and stability of the Higgs potential}
The equation \eqref{eq:mathieu} may lead to exponential growth of Higgs modes during preheating, destabilizing the low-energy vacuum. 
To see this, we restore dependence on $\lambda_h$ in \eqref{eq:mathieu}, such that the effective mass of the Higgs boson is $m_{h_{\text{eff}}}^2 = g \Phi^2 \cos^2(m_\phi t) + 3 \lambda_h \langle h^2 \rangle$.
We adopt a model in which the quartic coupling $\lh(\tilde{\mu})$ becomes negative at scales of  $\tilde{\mu} \geq h_{\rm inst} = 10^{10} \,\GeV$,~\footnote{As supported by the vast literature of Higgs vacuum stability, including \cite{Hung:1979dn,Degrassi:2012ry,Buttazzo:2013uya,Andreassen:2014gha}. This estimate holds approximately for the SM, for the currently measured value of the top mass $m_t = 173.0\pm0.4~{\rm GeV}$~\cite{Khachatryan:2015hba,Aaboud:2016igd,TevatronElectroweakWorkingGroup:2016lid,Tanabashi:2018oca} and strong coupling constant, $\alpha_s(m_Z) = 0.1181\pm0.0011$~\cite{Tanabashi:2018oca}.}
where $\tilde{\mu}$ is the renormalization scale, $\tilde\mu \sim \sqrt{\langle h^2 \rangle}$.
The exponential growth of Higgs modes pushes the Higgs variance \eqref{eq:higgsvariance} above the instability scale within a few semi-oscillations \cite{Enqvist:2016mqj}.
Above this scale, then, the effective Higgs potential is,
\beq
V_{\rm eff}(h) \approx \dfrac{1}{2} m^{2}_{h_{\text{eff}}} h^{2} - \dfrac{|\lh(\tilde\mu)|}{4}h^{4}
\eeq
such that there are vacua for $h=0$ and at infinity, with a barrier at 
\beq
h_{c} = \sqrt{-\dfrac{m^{2}_{h_{\text{eff}}}}{\lh(\tilde\mu)}} \sim \sqrt{\dfrac{g}{2 |\lh(\tilde\mu)|}} |\phi| \,.
\eeq
Consequently, if the Higgs field fluctuates beyond the critical value $h_{c}$ it will roll down to the catastrophic vacuum. 

The vacuum remains metastable during preheating if the parametric resonance ends before the vacuum decays.
Because $\phi$ is an oscillating function, there is a time interval $\Delta t$ around zero crossing $\phi \sim \Phi (m_\phi \Delta t)$ for which the effective mass of the Higgs in \eqref{eq:mathieu} is dominated by the negative self-interaction $3\lambda _h \langle h^2 \rangle=\delta m^{2}_{h}$.
This leads to a tachyonic amplification of the Higgs field, which lasts as long as the contribution from the negative self-interaction dominates,
\beq \label{eq:deltat}
\Delta t = \sqrt{\dfrac{6 \lh(\tilde\mu)\langle h^{2} \rangle }{g \Phi^{2} m_\phi^{2}}} \,.
\eeq
The vacuum decay time can be found by the condition that the growth rate during this interval is bounded by unity, which can be estimated as $\delta m_{h}^2 \Delta t = 3\lambda _h \langle h^2 \rangle \Delta t < 1$ \cite{Ema:2016kpf}. Using Eq.~\eqref{eq:deltat} with the Higgs variance~\eqref{eq:higgsvariance}, we find
\beq
t_{\text{dec}} \sim \dfrac{1}{2 \mu m_\phi} \ln{\left( \dfrac{16 \pi^{3/2}}{3 |\lh(\tilde\mu)|}\right)} + \dfrac{y^{2}_{t}\sqrt{3 g}}{32 \pi^{3/2} \mu}\left( \dfrac{M_{\rm Pl}}{m_\phi}\right)\,,
\eeq
where the second term accounts for perturbative decays into top quarks, and $\mu$ is the Floquet exponent (as before, $\mu \approx (\ln3)/2\pi$).
Preheating ends around $q\approx 1$, such that
\beq
t_{\text{res}} = \sqrt{\dfrac{g}{6 \pi}} \dfrac{M_{\rm Pl}}{m_\phi^{2}}\,.
\eeq
Ensuring that $t_{\text{res}}<t_{\text{dec}}$ implies a the stability condition given by
\beq \label{eq:gbound}
g< \dfrac{6 \pi}{4 \mu^{2}} \left[ \dfrac{m_\phi}{M_{\rm Pl}} \ln{\left( \dfrac{16 \pi^{3/2}}{3 |\lh(\tilde\mu)|}\right)} \right]^{2} \times \left(1 - \dfrac{y^{2}_{t}}{96 \sqrt{2} \pi^{2} \mu} \right)^{-2}\,.
\eeq
Using  $m_\phi= 1.3 \times 10^{-6} M_{\rm Pl}$ in the above, we find an upper bound $g<2 \times 10^{-8}$.
Note that we have derived this upper bound in the absence of the trilinear coupling; it is therefore a conservative estimate. 

Now we will consider the case of the trilinear coupling, $p\neq 0$.
This term generates an extra contribution to the Higgs effective mass, which is proportional to $\phi(t)$ and can induce a tachyonic resonance during the semi-oscillation for which $\phi(t)<0$ \cite{Dufaux:2006ee}. Notice that the time for the vacuum to decay that we calculated above only depends on the self-coupling of the Higgs $\lh$. Therefore, to a first approximation, we can find an upper bound on $\sigma$ using a similar calculation as above, 
\beq \label{eq:sigmabound}
\sigma < \dfrac{\sqrt{3 \pi}}{4 \mu^{*}}\left( \dfrac{m_\phi^{2}}{M_{\rm Pl}}\right) \ln \left( \dfrac{16 \pi^{3/2}}{3 |\lh(\tilde\mu)|}\right)
\eeq
where $\mu^{*}$ is the effective Floquet exponent for the Whittaker-Hill equation, which takes values of $~\mathcal{O}(0.1-1)$ in the instability regime \cite{Enqvist:2016mqj}. Using $m_\phi=1.3 \times 10^{-6} \, M_{\rm Pl}$ we find the upper bounds $\sigma \lesssim 5\times 10^{8} \, \GeV$ and $g\lesssim 2 \times 10^{-8}$.

We find numerically that even when saturating the upper bounds on the couplings $g$ and $\sigma$, preheating is not very efficient, and most of the energy density is still stored in the inflaton field when parametric resonance ends. Preheating into the sterile neutrino fields is also not expected to be efficient, due to Pauli-blocking \cite{Figueroa:2016wxr}.
Therefore perturbative decays are expected to take over and drain the energy density from the inflaton field.

\section{Stability constraints on leptogenesis after (p)reheating}\label{sec:results}
\begin{figure}
    \centering
    \includegraphics[width= .45\textwidth]{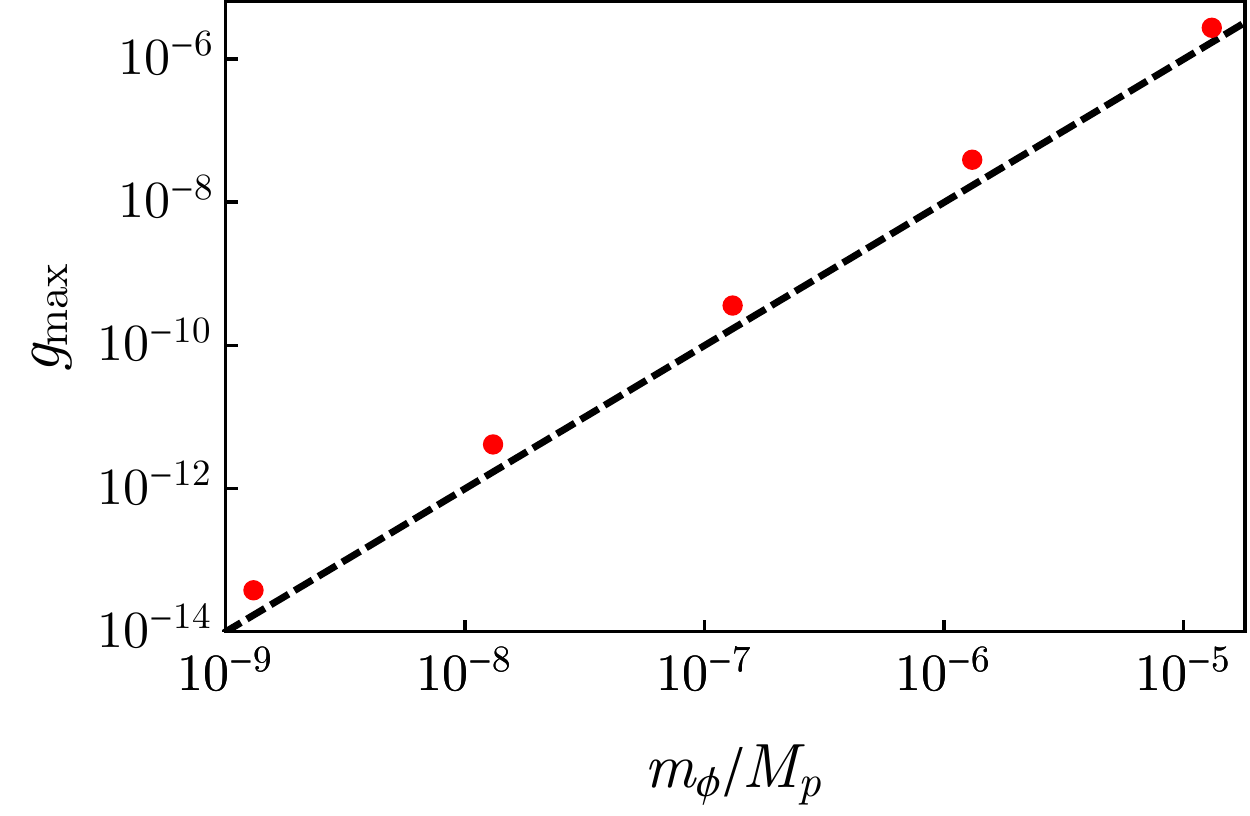}
    \includegraphics[width= .45\textwidth]{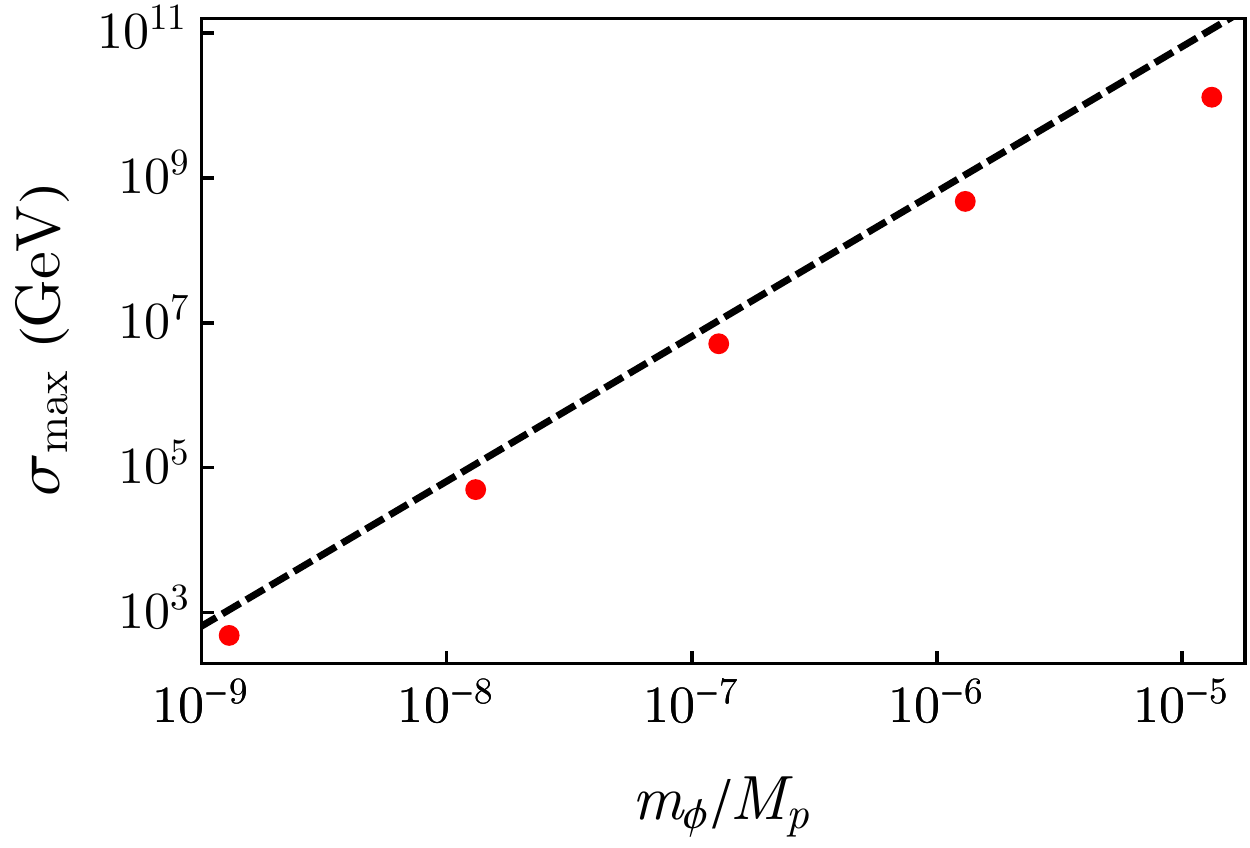}
    \caption{Upper bounds on the portal coupling $g$ and $\sigma$ respectively, found numerically (red points) and analytic estimates \eqref{eq:gbound} and \eqref{eq:sigmabound} (dashed lines).}
    \label{fig:gbound}
\end{figure}

We will now study constraints on leptogenesis from stability of the Higgs potential during preheating. 
We use \verb|LATTICEEASY|~\cite{Felder:2000hq} to find numerical upper bounds on the portal couplings as a function of $m_\phi/M_{p}$, using the condition that the instability timescale is longer than the parametric resonance regime lasts as described in the previous section. As in~\cite{Enqvist:2016mqj}, we use a simple estimate for the Higgs self-interaction $\lambda_h = - 10^{-2}$ at high scales, because we expect $\langle h^2 \rangle > 10^{10}$ GeV within the first semi-oscillations - we verify numerically that this assumption is valid. We also make the simplifying assumption that the bounds on the portal couplings are independent of each other.\footnote{This constitutes a conservative scenario: in reality, the constraints may be stronger.} The numerical upper bounds are shown in Fig.~\ref{fig:gbound} along with the approximate expressions of \eqref{eq:gbound} and \eqref{eq:sigmabound}.

The one-loop expressions \eqref{eq:portals} give rough lower bounds on the magnitudes of the portal couplings in a theory that is not finely tuned. Upper bounds on these couplings come from stability constraints, implying
\begin{equation}
\begin{aligned}
\left|\delta\sigma\right|&\simeq \frac{\lambda}{2\pi^2}(Y^\dag Y)_{11} M_1\log\frac{M_{\rm Pl}}{m_\phi}\lesssim \left|\sigma\right|<\sigma_{\rm max}\left(\frac{m_\phi}{M_{\rm Pl}} \right),
\\
\left|\delta g\right|&\simeq \frac{\lambda^2}{2\pi^2}(Y^\dag Y)_{11} \log\frac{M_{\rm Pl}}{m_\phi}\lesssim \left|g \right|<g_{\rm max}\left(\frac{m_\phi}{M_{\rm Pl}} \right),
\label{eq:portallimit}
\end{aligned}
\end{equation}
The upper bounds on the RHS of these equations are numerically found for different inflaton masses, and roughly scale as described in the previous section.
In particular, as was found in \eqref{eq:gbound}, the upper bound on the quartic portal coupling $g$ scales approximately with the square of the inflaton mass. The bound on the trilinear portal coupling is more uncertain up to order $\mathcal{O}(1)$ corrections because its dependency on the sign of $\sigma$ in an expanding Universe.
As we focus on the most conservative regime, in which the inflaton only couples to the lightest sterile neutrino, the constraints \eqref{eq:portallimit} are a function of the coupling $\lambda$, the inflaton mass, and the strongly correlated Yukawa coupling $(Y^\dagger Y)_{11}$ and sterile neutrino mass $M_1$ only.

\begin{figure}
    \centering
    \includegraphics[width=0.48\textwidth]{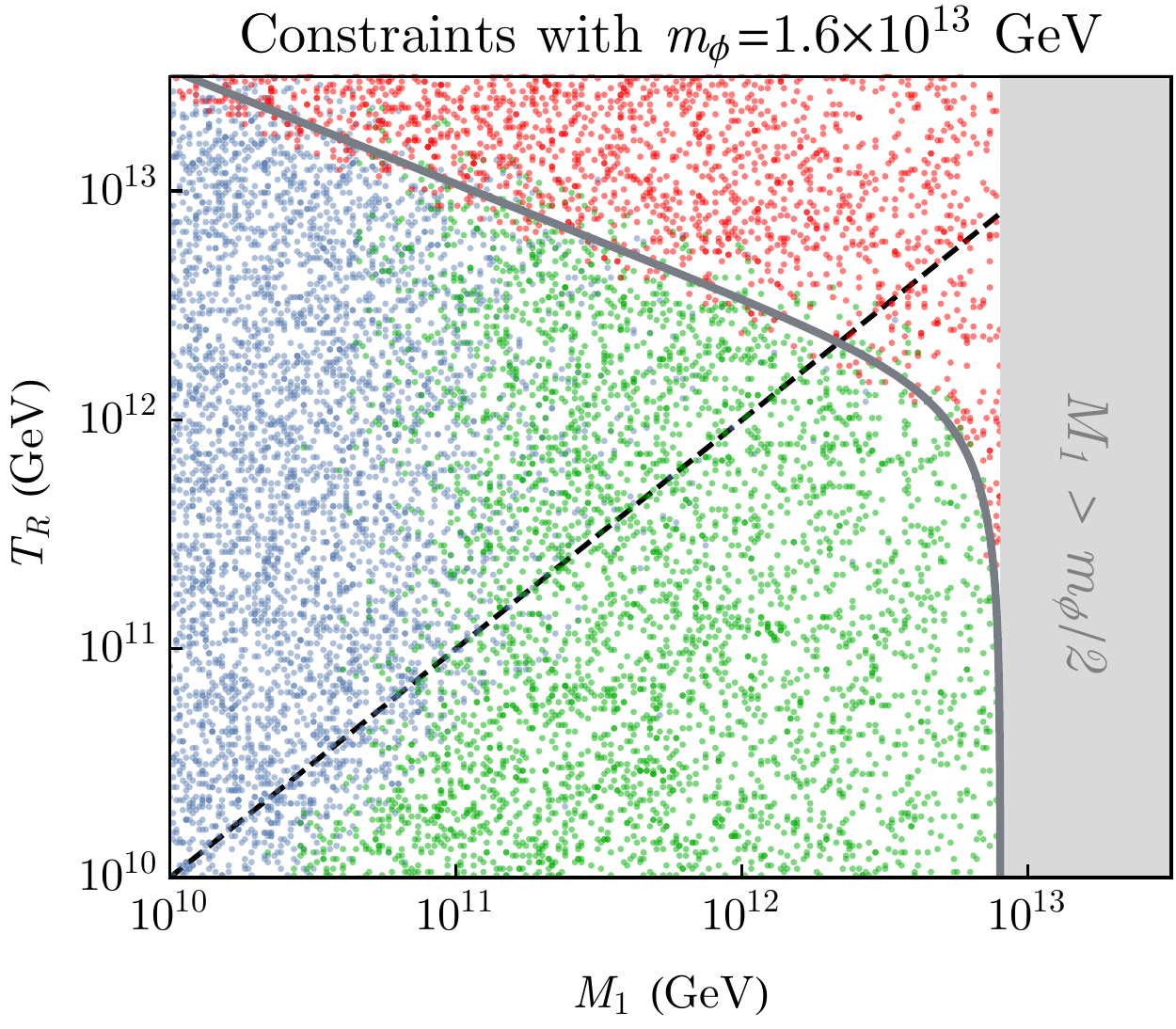}
    \includegraphics[width=0.48\textwidth]{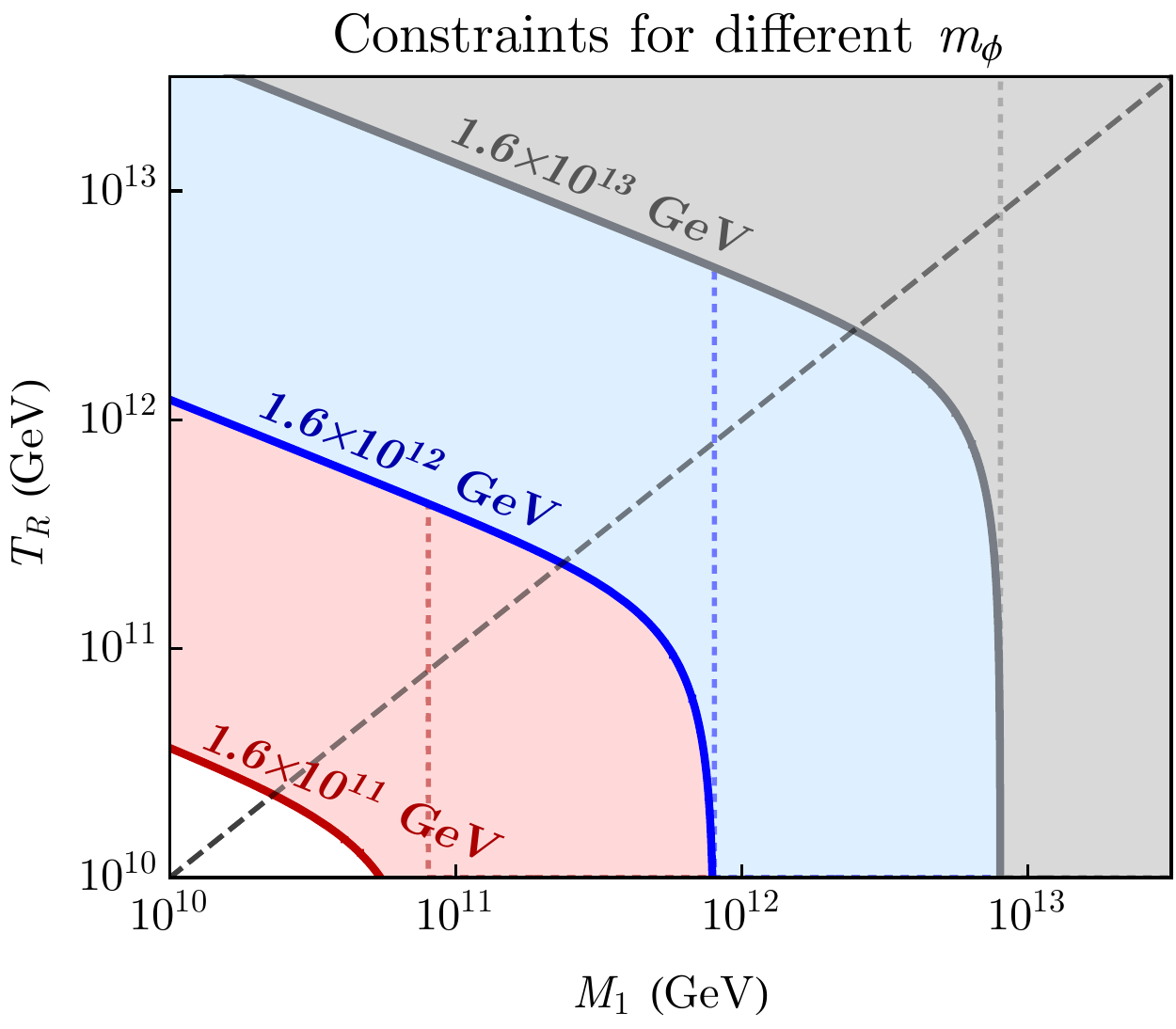}
    \caption{The $T_R$ vs. $M_1$ parameter space in the conservative scenario $\tilde\lambda={\rm diag}(\lambda,0)$. {\it Left panel}: the inflaton mass has been fixed to be $m_\phi = 1.3 \times 10^{-6} \, M_\text{Pl} = 1.6 \times 10^{13}$ GeV. All points reproduce the correct values of the light neutrino masses and mixing angles. We have imposed the zero temperature Higgs potential stability criterion of Ref.~\cite{Ipek:2018sai}. Green points correspond to a baryon asymmetry above our threshold~\eqref{eq:baryonass}; blue points result in an asymmetry below this. The red points are excluded by the preheating stability constraints of Eq.~\eqref{eq:portallimit}. The grey curve shows our semi-analytic estimate of the upper bound on $T_R$ Eq.~\eqref{eq:analyticconservative}. {\it Right panel}: The upper limit on $T_R$ from Eq.~\eqref{eq:analyticconservative}, varying the inflaton mass and using the bound on $g$ from Fig.~\ref{fig:gbound}.}
    \label{fig:Conservativeplot}
\end{figure}

In our analysis, we solve the Boltzmann equations in Section \ref{sec:leptogenesis} to find the baryon asymmetry,  using the full expressions for the decay rate, washout and CP violating parameter, $\epsilon _1$. We fix the inflaton mass at $m_\phi = 1.6 \times 10^{13}~{\rm GeV}$ (the mass of the inflaton in the chaotic inflation model) for clarity.\footnote{We consider this to be a conservative estimate, as most models of inflation have a lower inflaton mass~\cite{Akrami:2018odb}, for which the model will be more strongly constrained (cf. Fig. \ref{fig:Conservativeplot}).}  The complex angle $\sin z$ is parametrized as $\sin \phi _z e^{i \theta _z}$, and we allow $\phi _z$ and $\theta _z$ to vary over a $2 \pi$ range. We allow the other parameters to vary logarithmically, in the respective ranges $1\leq\lambda\leq10^{-8}$, $10^{10}\leq M_1/{\rm GeV}\leq 8\times 10^{12}~{\rm GeV}$, and $\sqrt{10}\leq M_2/M_1\leq 10^4$. The PMNS mixing angles are fixed to their central values and we allow unconstrained phases and angles to vary from 0 to $2\pi$ (or $4\pi$ for a Majorana phase).  
A point is considered to successfully reproduce the observed baryon asymmetry if $\eta_B$ is above the $3\sigma$ lower bound \cite{Cooke:2013cba,Ade:2015xua}
\begin{equation}\label{eq:baryonass}
    \eta _B -3 \Delta \eta _B = 5.45\times 10^{-10} \ .
\end{equation}
Creating too large an asymmetry is not a concern, as a reduction of an otherwise unobservable phase in the neutrino Yukawa matrix can lower the produced asymmetry.

We are interested in the physical scales $M_1$ and the reheating temperature $T_R$.
The latter is in this scenario well approximated by 
\begin{equation}
    T_R\approx \left( \frac{45}{4g_*\pi^3} \right)^{1/4} \sqrt{ M_{\rm Pl} \Gamma_{\phi\to N_1N_1}}\simeq\frac{\lambda}{4\pi}\left( \frac{45}{g_*\pi} \right)^{1/4} \sqrt{M_{\rm Pl} m_\phi}\left(1-\frac{4M_1^2}{m_\phi^2}\right)^{3/4}
\end{equation}
where $\Gamma_{\phi\to N_1N_1}$ is the two-body decay rate of the inflaton into the lightest sterile neutrino from~\eqref{eq:phidecayrate}. 
Note that sterile neutrino masses $M_1 > m_\phi/2 $ kinematically forbid perturbative decay to $N_1 N_1$. Now, using~\eqref{eq:YY11} and taking $\left|\sin z\right|^2\sim\left|\cos z\right|^2\sim 1/2$ in the absence of fine tuning, we can relate $(Y^\dag Y)_{11}$ to $M_1$ via
\begin{equation}
  (Y^\dag Y)_{11}\sim \frac{M_1m_3}{2v^2}\simeq 10^{-4}\left(\frac{M_1}{10^{11}~\rm GeV}\right).
  \label{eq:YY11vsM1}
\end{equation}
Using this and the upper bound on the $\phi^2\left|H\right|^2$ coupling, $g$, in~\eqref{eq:portallimit} we can find an upper bound on the reheat temperature of
\begin{equation}
\begin{aligned}
  T_R &\lesssim \left(\frac{45}{16g_\ast\pi}\right)^{1/4}\left(1-\frac{4M_1^2}{m_\phi^2}\right)^{3/4}\sqrt{\frac{g_{\rm max}}{\log M_{\rm Pl} / m_\phi}\frac{M_{\rm Pl}m_\phi}{M_1 m_3}}\,v
\\
&\simeq 1.1\times 10^{13}~{\rm GeV}\left(\frac{106.75}{g_\ast}\right)^{1/4}\left(\frac{g_{\rm max}}{2\times 10^{-8}}\right)^{1/2}\left(\frac{14}{\log M_{\rm Pl}/m_\phi}\right)^{1/2}
\\
&\quad\times\left(\frac{10^{11}~\rm GeV}{M_1}\right)^{1/2}\left(\frac{m_\phi}{1.6\times 10^{13}~\rm GeV}\right)^{1/2}\left(1-\frac{4M_1^2}{m_\phi^2}\right)^{3/4}.
\end{aligned}\label{eq:analyticconservative}
\end{equation}
Similarly, we can find an upper bound on $T_R$ from the $\phi \left|H\right|^2$ coupling, $\sigma$, in~\eqref{eq:portallimit},
\begin{equation}
\begin{aligned}
  T_R &\lesssim  4.7\times 10^{16}~{\rm GeV}\left(\frac{106.75}{g_\ast}\right)^{1/4}\left(\frac{\sigma_{\rm max}}{5\times 10^{8}~{\rm GeV}}\right)\left(\frac{14}{\log M_{\rm Pl}/m_\phi}\right)
\\
&\quad\times\left(\frac{10^{11}~\rm GeV}{M_1}\right)^{2}\left(\frac{m_\phi}{1.6\times 10^{13}~\rm GeV}\right)^{1/2}\left(1-\frac{4M_1^2}{m_\phi^2}\right)^{3/4},
\end{aligned}\label{eq:analyticconservative2}
\end{equation}
which is weaker than that in Eq.~(\ref{eq:analyticconservative}) over the parameter space we consider.

If we allow for a coupling of the inflaton to the heavier sterile neutrino, deviating from our choice $\tilde{\lambda} = \text{diag}(\lambda,0)$ these constraints become tighter; if the decay $\phi\to N_2 N_2$ is kinematically allowed, then the bound in (\ref{eq:analyticconservative}) is stronger by the factor $\sqrt{M_1/M_2}$ and that in (\ref{eq:analyticconservative2}) by $(M_1/M_2)^2$. If $\phi\to N_2 N_2$ is kinematically forbidden then the bounds in (\ref{eq:analyticconservative}) and (\ref{eq:analyticconservative2}) are both further strengthened beyond the above estimates by an additional factor of $M_1/M_2$.

Our results are shown in the $(T_R,M_1)$ parameter plane in the left panel of Fig.~\ref{fig:Conservativeplot} for $m_\phi=1.6\times10^{13}~{\rm GeV}$ assuming (conservatively) that the inflaton couples only to $N_1$, $\tilde{\lambda} = \text{diag}(\lambda,0)$. The dashed line at $T_R=M_1$ distinguishes between the regimes in which the sterile neutrino is produced in thermal equilibrium (roughly above the dashed line) and out of thermal equilibrium (roughly below the dashed line). The green points satisfy our criterion for the observed baryon asymmetry while the blue points lead to too small an asymmetry. The red points are ruled out by stability of the Higgs field during preheating, as they violate the conditions in~\eqref{eq:portallimit} (driven here by the upper bound on $g$). All points shown are required to satisfy the zero temperature Higgs stability constraints of~\cite{Ipek:2018sai}---this causes the density of points to decrease with increasing $M_1$. We also show the upper bound on $T_R$ from our semi-analytic estimate in Eq.~\eqref{eq:analyticconservative} as a solid curve. We see that this gives a good estimate of the boundary of the red points which fail the preheating stability criteria, with deviations from our estimate coming from the relationship between $(Y^\dagger Y)_{11}$ and $M_1$ in Eq.~(\ref{eq:YY11vsM1}) not being exact. 

To get a sense for how the bounds change for different inflaton masses, the right panel of Fig.~\ref{fig:Conservativeplot} shows our semi-analytic estimate of the upper bound on $T_R$ as a function of $M_1$ for $m_\phi=1.6\times10^{11}~{\rm GeV}$, $1.6\times10^{12}~{\rm GeV}$, and $1.6\times10^{13}~{\rm GeV}$, taking into account the dependence of $g_{\rm max}$ on $m_\phi$ as shown in Fig.~\ref{fig:gbound}. 
It is seen that the preheating stability constraint rules out much of the thermal leptogenesis parameter space, especially as the inflaton mass is decreased.

Deviating from our conservative scenario in which the inflaton dominantly couples to the lightest sterile neutrino would strengthen the bound on $T_R$ as a function of $M_1$, since the stability constraints would be driven by the larger $(Y^\dagger Y)_{22}$. The upper limits in~\eqref{eq:analyticconservative} and~\eqref{eq:analyticconservative2} would decrease by factors of $\sqrt{M_1/M_2}$ and $(M_1/M_2)^2$, respectively, along with a possible additional factor of $\lambda_{11}/\lambda_{22}$ depending on whether $\phi\to N_2N_2$ is kinematically allowed. Depending on the values of $M_1$ and $M_2$, this could cause the constraint on $\sigma$ to become more important than that on $g$.

\section{Discussion}\label{sec:discussion}
We have proposed a new cosmological constraint on vanilla leptogenesis. This constraint fills an important gap, as experimental access outside of the resonant regime is very limited. In simple models, the observed baryon asymmetry requires the lightest sterile neutrino to be roughly seven orders of magnitude above the weak scale, or heavier. 

In the vanilla leptogenesis scenario we study, the BAU is produced through the decay of heavy sterile neutrinos, which are reheat products of the inflaton. As Pauli-blocking renders parametric resonance into fermions inefficient, and preheating into Higgs bosons is severely constrained by stability considerations, the dominant reheat process is the perturbative decay into sterile neutrinos.
The available parameter space in the scenario we considered depends on the inflaton mass, as we assume that the two-body decay is to be kinematically allowed to allow for the production and (out-of-equilibrium) decay of the sterile neutrinos.

In this work we have assumed an instability scale of $\mu \sim 10^{10}$ GeV in the Higgs potential, although the results do not strongly depend on the precise value of this scale. 
We find that an instability scale of this order leads to tight constraints, in particular in the thermal leptogenesis parameter space. Non-thermal leptogenesis, which has received much interest due to the fact that it allows a lower reheating temperature (easing the tension with gravitino overproduction in models of supersymmetry \cite{Asaka:2000zh}), is also more compelling in the model we consider. 
 
The results in this paper imply that future precision measurements of the top and Higgs masses can shed further light on vanilla leptogenesis. Furthermore, future precision measurements of inflationary observables such as further constraints on the value of the spectral tilt as well as limits or the measurement of tensor modes in the CMB, may help to pinpoint the effective inflaton mass during preheating in particular inflation models. Such results will affect the severity of the constraints arising from stability.

The constraints in this paper apply to a sterile neutrino sector which couples directly to the inflaton. Non-minimal models may have a sterile sector as a product of a chain of decays. In such a case, the Higgs coupling to the inflaton may only arise at higher loop order, and the analysis in this paper should be modified for the relevant couplings \eqref{eq:portals}.

Stability constraints during pre-heating may be delayed if the Higgs has a non-minimal coupling to gravity \cite{Espinosa:2015qea}. Moreover, the Higgs vacuum itself may be stabilized in supersymmmetric models. Of course, the analysis can be adapted to include other catastrophic vacua on the SUSY scalar manifold. 
In this case, an interesting application of the analysis in this paper is in SUSY models which realize leptogenesis through an $N^3$ term in the superpotential \cite{Croon:2013ana,Ellis:2016ipm}. 
For such an analysis, one should be careful to include all contributions to the portal couplings; for example, there may be further contributions from Higgsino diagrams.

\acknowledgments
The authors would like to thank Gary Felder, Dani Figueroa, Jessica Turner, Lauren Pearce, David Morrissey, and Pasquale di Bari for helpful discussions. DC thanks the Lorentz Institute for hospitality during this work. NF is pleased to thank Bruno Villasenor for his help with the Julia implementation of the numerical solver for the Boltzmann equations. NF is partly supported by the U.S. Department of Energy grant number de{-}sc0010107. This work was performed in part at the Aspen Center for Physics, which is supported by National Science Foundation grant PHY-1607611. TRIUMF receives federal funding via a contribution agreement with the National Research Council of Canada and the Natural Science and Engineering Research Council of Canada.

\bibliography{references}
\end{document}